\documentclass[prl,aps]{revtex4}%
\usepackage{amssymb}
\usepackage{graphicx}
\usepackage{amsmath}
\usepackage{amsfonts}%
\begin{document}
%\twocolumn[\hsize\textwidth\columnwidth\hsize\csname@twocolumnfalse\endcsname
\title{Generalized Quantum Games with Nash Equilibrium }
\author{X.F.Liu$^{1,2}$, C.P.Sun$^{2,a,b}$}
\address{$^1$Department of Mathematics, Peking University, Beijing,
100871,China\\
$^2$Institute of Theoretical Physics, Chinese Academy of Science,
Beijing, 100080, China}
\begin{abstract}
We define generalized quantum games by  introducing  the coherent
payoff operators and propose a simple scheme to illustrate it. The
scheme is implemented  with a single spin qubit system and two
entangled qubit system. The Nash Equilibrium Theorem is proved for
the models. 
\end{abstract}
\maketitle]

\section{Introduction}

As a new vehicle to explore the exotic natures of quantum
information\cite{qinform}, quantum games were proposed as the
\textquotedblright quantization\textquotedblright\ of the classical
games\cite{Meyer,EWL99,qgame1}. Most recently a quantum game was implemented
via nuclear magnetic resonance (NMR) system\cite{exp-du}. It was demonstrated
that neither of the two players would win the game if they play rationally,
but if they adopt quantum strategies both of them would win. A classical game
\cite{cgame}, as is well known, consists of three elements - the players, the
strategies taken by the players and the payoff functions. In a gambling
process, a player takes a strategy without knowing the strategy adopted by the
other players. All players having taken actions simultaneously or
successively, the payoff is awarded to each player according to the payoff
function, which depends on the actions of all of the players. If a player can
maximize his payoff, one says he wins the game.

The early "quantization" of classical game is to replace the classical
strategy space with the quantum one consisting of unitary operations on a
quantum state. In the ordinary two player quantum game, one takes as the
initial quantum state an entangled state (EPR state \cite{EPR}correlating two
qubits in a distance and the players can take as strategies local unitary
operations acting on the two qubits separately. The payoff functions are
defined by the probabilities, rather than the probability amplitudes, of
projecting the final state to some chosen states. For this reason, we think
this definition of quantum game is incoherent. As in this "semi-classical game
theory ", the payoff function is based on the classical probability to a large
extent, there is a fundamental interest in generalizing it to a "fully-quantum
game theory ".

In this note we make two generalizations. First, we introduce the coherent
payoff functions in terms of certain probability amplitudes. Second, we loosen
the requirement that the initial state be an entangled qubit state. With these
generalizations, we can still prove the Nash Equilibrium Theorem
\cite{Nash}for a special type of quantum games, including single qubit case
and two entangled qubit case. Notice that the existence of Nash equilibrium is
an essential element in defining an interesting game. But it seems that this
point has not been fully realized. Indeed, in some proposed quantum games,
while new features have been demonstrated, Nash equilibrium appears to occur
accidentally for specifically-chosen parameters. A most recent work concerns
the universality of the Nash theorem \cite{NT}, but it requires a very large
quantum strategy space that consists of both unitary transformations and
non-unitary \textquotedblright quantum measurement \textquotedblright%
\ operations. Naturally, Nash equilibrium is more likely to occur in a bigger
strategy space. However, from the practical point of view, it is desirable to
introduce a reasonable quantum game in a limited strategy space with Nash
equilibrium whose existence is not the result of carefully choosing
parameters. Anyway, it should be possible to implement by practically physical processes.

\section{Generalized Quantum Games}

Let us describe the scheme of our generalized $N$ player quantum game in a
general framework. Fix a vector space $V$. Then mathematically the game is a
triple $\left(  S,\rho,P\right)  $ where $\rho\in End(V)$ being a density
matrix$,$ $S=\left(  S_{1},S_{2},\cdots,S_{N}\right)  ,S_{i}\subset End(V)$
consisting of unitary operators, and $P=(P_{1,}P_{2,}...,P_{N})$, $P_{i}$ $\in
End(V)$ being an Hermitian operator, called a payoff operator. For a given
quantum state described by a density matrix $\rho$ $,$the players transform it
by operations $U_{i}\in S_{i}$ $\left(  i=1,2,\cdots,N\right)  $
simultaneously for a static type game or in succession for a dynamic type
game. Based on the obtained final state
\begin{equation}
\rho_{f}=(%
%TCIMACRO{\dprod \limits_{k=1}^{N}}%
%BeginExpansion
{\displaystyle\prod\limits_{k=1}^{N}}
%EndExpansion
U_{k})\rho(%
%TCIMACRO{\dprod \limits_{k=1}^{N}}%
%BeginExpansion
{\displaystyle\prod\limits_{k=1}^{N}}
%EndExpansion
U_{k})^{\dagger}%
\end{equation}
the payoff $f_{i}$ for the i'th player is calculated according to the formula
\begin{equation}
f_{i}=Tr(P_{i}\rho_{f})\in R
\end{equation}
Notice that in this generalized version, the operations $U_{k}$ are not
required to be localized and so $[U_{k,}U_{s}]=0$ $(s\neq k)$is not
\ necessarily true. When $[\rho_{f},P_{i}]\neq0$ or $[U_{k},P_{i}]\neq0,$the
role of the off-diagonal elements in $\rho_{f}$ and $P_{k}$ reflect the
quantum coherence character of the payoff function. Notice that quantum
coherence plays crucial role in quantum computing and quantum information, but
it is sensitive to quantum meaurement \cite{sun-qm} and any
evironment-couplings with it \cite{sun-dec}.

The above described $N$ player game includes \ the original quantum game as
\ a special case. In that case, an N-multiple entangled pure state
$|\sigma\rangle\in V=V_{1}\otimes V_{2}\otimes...\otimes V_{N}$ is used as an
initial state. Here, $V_{k}$ is the space under the action of the operation
$U_{k}$ by the k'th player. One necessarily has $[U_{k,}U_{s}]=0$ and
$|\sigma\rangle$ is transformed by independent operations $U_{k}$
simultaneously in different spatial locations. The payoff
\begin{equation}
f_{k}(\sigma)=\sum_{j_{1},j_{2},...,j}C_{j_{1},j_{2},...,j_{2}}^{[k]}%
|\langle\sigma_{j_{1}}\sigma_{j_{2},}...\sigma_{j_{N}}|E\rangle|^{2}%
\end{equation}
for each player is calculated based on the probabilities $|\langle
\sigma_{j_{1}}\sigma_{j_{2},}...\sigma_{j_{N}}|E\rangle|^{2}$ for projections
of final state $|E\rangle=U_{1}U_{2,}...U_{N}|\sigma\rangle$ onto the basis
$|\sigma_{j_{1}}\sigma_{j_{2},}...\sigma_{j_{N}}\rangle.$Here, $C_{j_{1}%
,j_{2},...,j_{2}}^{[k]}$ are the real parameters assigned for a given game.
Obviously, it is a special case of our generalized version with the payoff
matrix
\begin{equation}
P_{k}=\sum_{j_{1},j_{2},...,j}C_{j_{1},j_{2},...,j_{2}}^{[k]}|\sigma_{j_{1}%
}\sigma_{j_{2},}...\sigma_{j_{N}}\rangle\langle\sigma_{j_{1}}\sigma_{j_{2}%
,}...\sigma_{j_{N}}|
\end{equation}
and the initial density matrix $\rho=|\sigma\rangle\left\langle \sigma
\right\vert .$

Next we consider two models of the above generalized quantum game with single
qubit and two entangled qubits, in which the Nash theorem holds. For
convenience, we start from an abstract classical game called GAME A defined on
a subset of $S^{1}\times S^{1}:$ $\left\{  (\theta,\varphi)|0\leq
\theta,\varphi\leq\frac{\pi}{2}\right\}  $. We will show that the two
generalized games are mathematically equivalent to the GAME A. So we need only
to see whether the Nash theorem holds universally for the GAME A.

\section{\bigskip Classical Abstract Game with Nash equilibrium\textit{ }on
$S^{1}\times S^{1}$}

The GAME A is described as follows. The two players have the strategy spaces
$\left\{  \theta|0\leq\theta\leq\frac{\pi}{2}\right\}  \subset$ $S^{1}$ and
$\left\{  \varphi|0\leq\varphi\leq\frac{\pi}{2}\right\}  $ $\subset$ $S^{1}%
$respectively. The pay off function for the $i-th$ player can be written in
the form
\begin{equation}
f_{i}\left(  \theta,\varphi\right)  =p_{i}+q_{i}\sin\left(  \theta
+\varphi+\Psi_{i}\right)
\end{equation}
where $q_{i}>0$ and $\Psi_{i}\in\left[  -\frac{\pi}{2},\frac{\pi}{2}\right]
.$

\textit{Proposition. For \ Game A, there exists a Nash equilibrium.}

Proof. We imitate the proof of Nash Equilibrium Theorem. From the conditions
$0\leq\theta,\varphi\leq\frac{\pi}{2}$ and $\Psi_{1}\in\left[  -\frac{\pi}%
{2},\frac{\pi}{2}\right]  $ \ we observe that for a fixed $\varphi$ there
exists exactly one $\theta,$ which we \ denote by $\chi\left(  \varphi\right)
,$ that \ maximizes the pay off function $f_{1}.$ Similarly, for a \ fixed
$\theta$ there is exactly one $\varphi,$ which we denote by $\varkappa\left(
\theta\right)  ,$ that maximizes the pay off function $f_{2}.$ So we can
define a map \ $g$ from the convex set $\left[  0,\frac{\pi}{2}\right]
\times$ $\left[  0,\frac{\pi}{2}\right]  $ to \ itself such that
\begin{equation}
g\left(  \theta,\varphi\right)  =\left(  \chi\left(  \varphi\right)
,\varkappa\left(  \theta\right)  \right)  .
\end{equation}
It is easy to show that $g$ is a continuous map. Hence, by Brower Fixed Point
\ Theorem $g$ has a fixed point $\left(  \theta_{0},\varphi_{0}\right)  .$ It
is readily \ verified that $\left(  \theta_{0},\varphi_{0}\right)  $ is a Nash
equilibrium. The proof of the proposition is thus completed.

\bigskip The Nash equilibrium of GAME A can actually be calculated explicitly.
The results, depending on the values of $\Psi_{1}$ and $\Psi_{2}$, are as
follows.Suppose $\Psi_{1}=\Psi_{2}\ =\Psi.$ Then the Nash equilibrium might
not be unique. If $\Psi\in\left[  -\frac{\pi}{2},0\right]  $, then each point
$\left(  \theta_{0},\varphi_{0}\right)  $ in $\left[  -\Psi,\frac{\pi}%
{2}\right]  \times\left[  -\Psi,\frac{\pi}{2}\right]  $ satisfying $\theta
_{0}+\varphi_{0}+\Psi=\frac{\pi}{2}$ is a Nash equilibrium; If $\Psi\in\left[
0,\frac{\pi}{2}\right]  $, then each point $\left(  \theta_{0},\varphi
_{0}\right)  $ in $\left[  0,\frac{\pi}{2}-\Psi\right]  $ $\times\left[
0,\frac{\pi}{2}-\Psi\right]  $ satisfying $\theta_{0}+\varphi_{0}+\Psi
=\frac{\pi}{2}$ is a Nash equilibrium. Suppose $\Psi_{1}\neq\Psi_{2}.$ Then
the Nash equilibrium is unique. Precisely, \ there are\ the following possibilities.

1) $\Psi_{1},\Psi_{2}\in\left[  -\frac{\pi}{2},0\right]  ,\Psi_{1}>\Psi
_{2},\left(  \theta_{0},\varphi_{0}\right)  =\left(  -\Psi_{1},\frac{\pi}%
{2}\right)  ;$

2) $\Psi_{1},\Psi_{2}\in\left[  -\frac{\pi}{2},0\right]  ,\Psi_{1}<\Psi
_{2},\left(  \theta_{0},\varphi_{0}\right)  =\left(  \frac{\pi}{2},-\Psi
_{2}\right)  ;$

3) $\Psi_{1}\in\left[  -\frac{\pi}{2},0\right]  ,\Psi_{2}\in\left[
0,\frac{\pi}{2}\right]  ,\left(  \theta_{0},\varphi_{0}\right)  =\left(
\frac{\pi}{2},0\right)  ;$

4) $\Psi_{1}\in\left[  0,\frac{\pi}{2}\right]  ,\Psi_{2}\in\left[  -\frac{\pi
}{2},0\right]  ,\left(  \theta_{0},\varphi_{0}\right)  =\left(  0,\frac{\pi
}{2}\right)  ;$

5) $\Psi_{1},\Psi_{2}\in\left[  0,\frac{\pi}{2}\right]  ,\Psi_{1}>\Psi
_{2},\left(  \theta_{0},\varphi_{0}\right)  =\left(  0,\frac{\pi}{2}-\Psi
_{2}\right)  ;$

6) $\Psi_{1},\Psi_{2}\in\left[  0,\frac{\pi}{2}\right]  ,\Psi_{1}<\Psi
_{2},\left(  \theta_{0},\varphi_{0}\right)  =\left(  \frac{\pi}{2}-\Psi
_{1},0\right)  .$

It is easy to prove these results. \ For example, when $\Psi_{1},\Psi_{2}%
\in\left[  -\frac{\pi}{2},0\right]  ,$ we have
\begin{align}
\chi\left(  \varphi\right)   &  =\left\{
\begin{array}
[c]{ll}%
\frac{\pi}{2}, & -\frac{\pi}{2}\leq\varphi+\Psi_{1}\leq0;\\
\frac{\pi}{2}-(\varphi+\Psi_{1}), & 0\leq\varphi+\Psi_{1}\leq\frac{\pi}{2}.
\end{array}
\right.  \nonumber\\
\varkappa\left(  \theta\right)   &  =\left\{
\begin{array}
[c]{ll}%
\frac{\pi}{2}, & -\frac{\pi}{2}\leq\theta+\Psi_{2}\leq0;\\
\frac{\pi}{2}-(\theta+\Psi_{2}), & 0\leq\theta+\Psi_{2}\leq\frac{\pi}{2}.
\end{array}
\right.
\end{align}
If $\Psi_{1}>\Psi_{2},$ it then follows from the fixed point condition
$\left(  \chi\left(  \varphi_{0}\right)  ,\varkappa\left(  \theta_{0}\right)
\right)  =\left(  \theta_{0},\varphi_{0}\right)  $ that $\left(  \theta
_{0},\varphi_{0}\right)  =\left(  -\Psi_{1},\frac{\pi}{2}\right)  .$ The other
results can be proved in a similar way. We notice that when $\Psi_{1}=\Psi
_{2}\ ,$ if the two players do not exchange information, there is little
chance of reaching a Nash equilibrium. On the other hand, when $\Psi_{1}%
\neq\Psi_{2},$ as the Nash equilibrium is unique it will appear, but at the
equilibrium at most only one player will maxize the \ payoff \ function.

\section{One Qubit Realization}

Now we consider the first realization of Game A only using one qubit, which is
similar to the classical game of roulette. Given an initial state $\rho,$ the
first player operates on it by a rotation
\begin{equation}
U\left(  \theta\right)  =\left(
\begin{array}
[c]{cc}%
\cos\theta & -\sin\theta\\
\sin\theta & \cos\theta
\end{array}
\right)  ,
\end{equation}
and then the second player operates on the resulted state by another rotation
$U\left(  \varphi\right)  .$Since the two operations $U\left(  \theta\right)
$ and $U\left(  \varphi\right)  $ take action in succession, this is a game of
dynamic type. Assign two payoff matrices $P_{1}$ and $P_{2}$ to \ the two
players. By definition they are Hermitian operators. Generally, we can write
\begin{equation}
P_{i}=\left(
\begin{array}
[c]{cc}%
a_{i} & b_{i}\\
\overline{b_{i}} & d_{i}%
\end{array}
\right)  ,i=1,2
\end{equation}
with respect to the basis $\left\{  \left\vert 0\right\rangle ,\left\vert
1\right\rangle \right\}  ,$ where $a_{i},d_{i}$ are real numbers. The complex
elements $b_{i}$ characterize the quantum coherence of the pay off functions
$f_{i}$ $\left(  i=1,2\right)  $ for the two players
\begin{equation}
f_{i}=tr\left(  P_{i}U\left(  \theta+\varphi\right)  \rho U\left(
\theta+\varphi\right)  ^{\dag}\right)  .
\end{equation}
To acquire the non-trivial nature of quantum coherence it should be required
that $\left[  P_{1},P_{2}\right]  $ and $[P_{i},U\left(  \theta\right)
]\neq0.$Otherwise one\ can simultaneously diagonalize $P_{1},$ $P_{2}$ and
$U\left(  \theta\right)  $ and the defined game would be trivial.

First, we take as the initial state the pure state
\begin{equation}
\left\vert \psi\right\rangle =\frac{1}{\sqrt{2}}\left(  \left\vert
0\right\rangle +\left\vert 1\right\rangle \right)
\end{equation}
of a qubit. Then we have%
\begin{align}
U\left(  \theta\right)  \left\vert 0\right\rangle  &  =\cos\theta\left\vert
0\right\rangle +\sin\theta\left\vert 1\right\rangle ,\nonumber\\
U\left(  \theta\right)  \left\vert 1\right\rangle  &  =-\sin\theta\left\vert
0\right\rangle +\cos\theta\left\vert 1\right\rangle .
\end{align}
After simple calculation we obtain
\begin{align}
f_{i}  &  =\frac{1}{2}[\left(  a_{i}+d_{i}\right)  +\left(  a_{i}%
-d_{i}\right)  \sin2\left(  \theta+\varphi\right) \nonumber\\
&  +\left(  b_{i}+\overline{b_{i}}\right)  \cos2\left(  \theta+\varphi\right)
].
\end{align}
If $a_{i}-d_{i}\geq0$ then $f_{i}$ can be rewritten as
\begin{equation}
f_{i}=p_{i}+q_{i}\sin\left(  2\left(  \theta+\varphi\right)  +\Psi_{i}\right)
,
\end{equation}
where
\begin{align}
p_{i}  &  =\frac{1}{2}\left(  a_{i}+d_{i}\right)  ,\nonumber\\
q_{i}  &  =\frac{1}{2}\sqrt{\left(  a_{i}-d_{i}\right)  ^{2}+\left(
b_{i}+\overline{b_{i}}\right)  ^{2}},\nonumber\\
\Psi_{i}  &  =\arctan\frac{\left(  b_{i}+\overline{b_{i}}\right)  }{\left(
a_{i}-d_{i}\right)  }.
\end{align}
Clearly, if we put the \ restrictions $a_{i}-d_{i}\geq0$ and $0\leq
\theta,\varphi\leq\frac{\pi}{4}$ then this quantum game is a realization of
Game A. Thus in this case it has a Nash equilibrium.

Next, we take the mixed state%
\begin{equation}
\rho=\left(
\begin{array}
[c]{cc}%
p & 0\\
0 & 1-p
\end{array}
\right)  ,0\leqslant p\leqslant1
\end{equation}
as the initial state. In this case we have%
\begin{align*}
f_{i}  &  =\frac{1}{2}[\left(  a_{i}+d_{i}\right)  +\left(  1-2p\right)
\left(  b_{i}+\overline{b_{i}}\right)  \sin2\left(  \theta+\varphi\right) \\
&  +\left(  1-p\right)  \left(  d_{i}-a_{i}\right)  \cos2\left(
\theta+\varphi\right)  ].
\end{align*}
It does not harm to assume $p\leqslant\frac{1}{2}.$ Then if $b_{i}%
+\overline{b_{i}}\geqslant0,$ $f_{i}$ can be rewritten as
\[
f_{i}=p_{i}+q_{i}\sin\left(  2\left(  \theta+\varphi\right)  +\Psi_{i}\right)
,
\]
where%
\begin{align}
p_{i}  &  =\frac{1}{2}\left(  a_{i}+d_{i}\right)  ,\nonumber\\
q_{i}  &  =\frac{1}{2}\sqrt{\left(  d_{i}-a_{i}\right)  ^{2}\left(
1-p\right)  ^{2}+\left(  b_{i}+\overline{b_{i}}\right)  ^{2}\left(
1-2p\right)  ^{2}},\\
\Psi_{i}  &  =\arctan\frac{\left(  1-p\right)  \left(  d_{i}-a_{i}\right)
}{\left(  1-2p\right)  \left(  b_{i}+\overline{b_{i}}\right)  }.\nonumber
\end{align}
Thus this game with the additional restriction $0\leq\theta,\varphi\leq
\frac{\pi}{4}$ is also a realization of GAME A.

\bigskip

\section{Two Qubit Realization}

Our next model is a static quantum game with two qubits. We take the quantum
entangled state
\begin{equation}
\left\vert \phi\right\rangle =\frac{1}{\sqrt{2}}\left(  \left\vert
0\right\rangle \otimes\left\vert 1\right\rangle +\left\vert 1\right\rangle
\otimes\left\vert 0\right\rangle \right)
\end{equation}
for a simple two qubit system. The two players independently operate on the
first and the second qubits by the above defined $U\left(  \theta\right)  $
and $U\left(  \varphi\right)  $ respectively. Notice that, since the two local
operations $U_{1}\left(  \theta\right)  =U\left(  \theta\right)  \otimes1$ and
$U_{2}\left(  \varphi\right)  =1\otimes U\left(  \varphi\right)  $ can take
action simultaneously or in succession, we can realize this two-bit game both
in the static and dynamic ways. As in the first model, two Hermitian payoff
operators $P_{1}$ and $P_{2}$ are assigned to the two players and the two pay
off functions $f_{i}$ $\left(  i=1,2\right)  $ are defined as%
\begin{equation}
f_{i}=tr\left(  P_{i}\left(  U\left(  \theta\right)  \otimes U\left(
\varphi\right)  \right)  \left\vert \phi\right\rangle \left\langle
\phi\right\vert \left(  U\left(  \theta\right)  \otimes U\left(
\varphi\right)  \right)  ^{\dagger}\right)  .
\end{equation}
Suppose that $P_{i}$ has the matrix representation%
\begin{equation}
\left(  P_{i}\right)  _{kl}=x_{kl}^{i},x_{kl}^{i}=\overline{x_{lk}^{i}%
},k,l=1,2,3,4,
\end{equation}
with respect to the basis
\begin{equation}
\left\{  \left\vert 0\right\rangle \otimes\left\vert 0\right\rangle
,\left\vert 0\right\rangle \otimes\left\vert 1\right\rangle ,\left\vert
1\right\rangle \otimes\left\vert 0\right\rangle ,\left\vert 1\right\rangle
\otimes\left\vert 1\right\rangle \right\}  .
\end{equation}
Then by direct calculation we obtain
\begin{align}
4f_{i} &  =\left(  \sum_{j=1}^{4}x_{jj}^{i}+2\operatorname{Re}x_{23}%
^{i}-2\operatorname{Re}x_{14}^{i}\right)  \nonumber\\
&  +(-x_{11}^{i}+x_{22}^{i}+x_{33}^{i}-x_{44}^{i}\nonumber\\
&  +2\operatorname{Re}x_{23}^{i}+2\operatorname{Re}x_{14}^{i})\cos2\left(
\theta+\varphi\right)  \nonumber\\
&  -2(\operatorname{Re}x_{12}^{i}+\operatorname{Re}x_{13}^{i}%
+\operatorname{Re}x_{24}^{i}\nonumber\\
&  +\operatorname{Re}x_{34}^{i})\sin2\left(  \theta+\varphi\right)  .
\end{align}
If
\begin{equation}
\operatorname{Re}x_{12}^{i}+\operatorname{Re}x_{13}^{i}+\operatorname{Re}%
x_{24}^{i}+\operatorname{Re}x_{34}^{i}\leq0,
\end{equation}
the pay off functions can be rewritten as
\begin{equation}
f_{i}=p_{i}+q_{i}\sin\left(  2\left(  \theta+\varphi\right)  +\Psi_{i}\right)
,
\end{equation}
where
\begin{align}
p_{i} &  =\frac{1}{4}(\sum_{j=1}^{4}x_{jj}^{i}+2\operatorname{Re}x_{23}%
^{i}-2\operatorname{Re}x_{14}^{i}),\nonumber\\
q_{i} &  =[\frac{1}{4}(-x_{11}^{i}+x_{22}^{i}+x_{33}^{i}-x_{44}^{i}\nonumber\\
&  +2\operatorname{Re}x_{23}^{i}+2\operatorname{Re}x_{14}^{i})^{2}+\nonumber\\
&  4\left(  \operatorname{Re}x_{12}^{i}+\operatorname{Re}x_{13}^{i}%
+\operatorname{Re}x_{24}^{i}+\operatorname{Re}x_{34}^{i}\right)  ^{2}%
]^{1/2}\nonumber\\
\Psi_{i} &  =-\arctan\frac{A}{B}:\nonumber\\
A &  =-x_{11}^{i}+x_{22}^{i}+x_{33}^{i}-\nonumber\\
&  x_{44}^{i}+2\operatorname{Re}x_{23}^{i}+2\operatorname{Re}x_{14}%
^{i}\nonumber\\
B &  =\operatorname{Re}x_{12}^{i}+\operatorname{Re}x_{13}^{i}%
+\operatorname{Re}x_{24}^{i}+\operatorname{Re}x_{34}^{i}%
\end{align}
Thus we conclude that this quantum game with the restrictions
\begin{equation}
\operatorname{Re}x_{12}^{i}+\operatorname{Re}x_{13}^{i}+\operatorname{Re}%
x_{24}^{i}+\operatorname{Re}x_{34}^{i}\leq0
\end{equation}
and $0\leq\theta,\varphi\leq\frac{\pi}{4}$ also realizes Game A.

\section{Remarks}

In sum, we introduce in this paper a new type of quantum game and prove the
Nash Equilibrium Theorem. We also\ calculate the equilibrium explicitly. In
the two models though "coherent " pay-off functions of quantum nature are
introduced, the universal existence of Nash equilibrium follows from the
simple mathematical structure of the classical Game A. In this sense, the game
is not really quantized. On the other hand, this suggests the possibility of
studying quantum games from an abstract point of view, transcending concrete examples.

Finally we point out that the interesting examples introduced in this letter
can easily be implemented in experiments. For the above one-spin qubit model,
the operation $U\left(  \theta\right)  $ can be realized as a Rabi rotation of
angle $\theta$ around z-axe in a spin procession experiment. If we take the
two payoff matrices to be the Pauli matrices $\sigma_{z}$ and $\sigma_{x}$
respectively, then the measurement of the payoff function is to see the
average polarization of spin along $x-axe$ and $z-axe$. Thus this quantum game
is implementable by a qubit NMR quantum computing system\cite{NMR}, described
as one nucleolus spin in a magnetic field driven by a radio frequency (rf )
field. This system realizes a single qubit quantum logic gate.

\textit{This work is supported by the NSF of China and the knowledge
Innovation Program (KIP) of the Chinese Academy of Science. It is also founded
by the National Fundamental Research Program of China with No 001GB309310. We
also thank Jiangfeng Du for the helpful discussions with him.}

\end{document}